\documentclass[prd, twocolumn, nofootinbib, floatfix, a4paper]{revtex4}
\usepackage{amsmath}
\usepackage{graphicx}
\usepackage{dcolumn}
\usepackage{bm}
\usepackage{epsfig}
\usepackage{amssymb,latexsym,mathrsfs}
\usepackage{graphicx}
\usepackage{color}

\newcommand{\be}{\begin{equation}}
\newcommand{\ee}{\end{equation}}
\newcommand{\bea}{\begin{eqnarray}}
\newcommand{\eea}{\end{eqnarray}}

\newcommand{\cvis}{c_{\rm vis}}
\newcommand{\cs}{c_{\rm s}}

\newcommand{\od}{\Omega_{\rm de}}
\newcommand{\omegae}{\Omega_{\rm e}}
\newcommand{\ome}{\Omega_{\rm e}}
\newcommand{\om}{\Omega_m}

\newcommand{\neff}{N_{\rm {eff}}}
\newcommand{\neffnu}{N_{\rm {eff}}^{\nu}}

\newcommand{\aeq}{a_{\rm {eq}}}

\newcommand{\Mpc}{{\rm ~Mpc}}
\newcommand{\Gyr}{{\rm ~Gyr}}

\begin{document}
\title{Limits on Dark Radiation, Early Dark Energy, and Relativistic 
Degrees of Freedom}

\author{Erminia Calabrese$^1$,  Dragan Huterer$^2$, Eric V.\ Linder$^{3,4}$, 
Alessandro Melchiorri$^1$, Luca Pagano$^5$} 
\affiliation{$^1$ Physics Department and INFN, Universita' di Roma ``La Sapienza'', Ple Aldo Moro 2, 00185, Rome, Italy.}
\affiliation{$^2$ Department of Physics, University of Michigan, 450 Church St, Ann Arbor, MI 48109, USA.}
\affiliation{$^3$ Berkeley Lab \& University of California, Berkeley, CA 94720, USA.} 
\affiliation{$^4$ Institute for the Early Universe WCU, Ewha Womans University, Seoul, Korea} 
\affiliation{$^5$ Jet Propulsion Laboratory, California Institute of Technology, 4800 Oak Grove Drive, Pasadena CA 91109, USA.}

\begin{abstract}
Recent cosmological data analyses hint at the presence of an extra
relativistic energy component in the early universe.  This component
is often parametrized as an excess of the effective neutrino number
$\neff$ over the standard value of 3.046.  The excess relativistic
energy could be an indication for an extra (sterile) neutrino, but
early dark energy and barotropic dark energy also contribute to the
relativistic degrees of freedom.  We examine the capabilities of
current and future data to constrain and discriminate between these
explanations, and to detect the early dark energy density
associated with them.  We find that while early dark energy does not
alter the current constraints on $\neff$, a dark radiation component,
such as that provided by barotropic dark energy models, can
substantially change current constraints on $\neff$, bringing its
value back to agreement with the theoretical prediction.  Both dark energy models also
have implications for the primordial mass fraction of Helium $Y_p$ and
the scalar perturbation index $n_s$.  The ongoing Planck satellite
mission will be able to further discriminate between sterile
neutrinos and early dark energy.

\end{abstract}

\maketitle

\section{Introduction}
\label{sec:intro}

The precision of theoretical modelling and observational measurements of 
the cosmic microwave background (CMB) temperature and polarization 
anisotropies from satellites and ground based experiments 
\cite{wmap7, acbar, quad, act} 
has opened the exciting possibility of addressing key questions about the 
nature of dark energy, dark matter and primordial inflation. 
Interestingly, a first hint of some new physics may be showing up in the 
high redshift universe from CMB and Big Bang nucleosynthesis (BBN) data.  
Recent analyses (for example 
\cite{seljak, mangano, hamann, wmap7, act}) may be hinting at the need for an 
extra, dark, relativistic energy component. 

If further data confirms this, it could suggest new dark matter such
as a sterile neutrino \cite{hamann} or a decaying particle
\cite{Zhang:2007zzh, Chluba:2008aw, Fischler:2010xz}, or nonstandard thermal history
\cite{Bashinsky:2003tk}.  Or it could indicate that dark energy does
not fade away to the $\sim 10^{-9}$ fraction of the energy density at
CMB recombination that is predicted by the cosmological constant.
Indeed, some proposed particle physics explanations for dark energy
involve scaling fields \cite{Ferreira:1997hj,Liddle:1998xm} with an
early energy density which is a constant fraction of the energy
density of the dominant component.  Another possibility is that 
the evidence for the
extra relativistic component may be signaling the presence of a 
``dark radiation'' component in the early universe, as 
predicted by certain higher dimension braneworld
scenarios \cite{Binetruy:1999hy}.

Any of these would be exciting extensions to the standard, concordance
model.  Uncovering new degrees of freedom would be of great
importance, and distinguishing between the possible origins could give
valuable insight into physics and cosmology.  The extra, dark
contribution to the total relativistic energy density in the early
Universe is generally phrased in terms of the energy density of
neutrinos -- the known dark, (early) relativistic component.  One
defines the effective number of neutrinos $\neff$ through :
\begin{equation}
\rho_\nu=  \rho_\gamma\, \frac{7}{8}\, \left( \frac{4}{11} \right)^{4/3} \neff \ , 
\label{eq:neffdef} 
\end{equation} 
where $\rho_\nu$ is the neutrino energy density and $\rho_\gamma$ is
the CMB photon energy density, with value today
$\rho_{\gamma,0}\approx 4.8 \times 10^{-34}$ g\,cm$^{-3}$.  In the
standard model, with three massless neutrinos with zero chemical
potential and in the limit of instantaneous decoupling, $\neff =
3$. The inclusion of entropy transfer between neutrinos and the
thermal bath modifies this number to about $\neff =
3.046$ at the CMB epoch (see e.g.~\cite{mangano}).

The recent analysis of \cite{act} that combined CMB data with
measurements of baryon acoustic oscillations (BAO) and the Hubble
constant reported an excess of the relativistic neutrino number,
$\neff=4.6\pm0.8$ at $68 \%$ c.l., disfavoring the standard value at
about two standard deviations.  This is compatible with previous
analyses \cite{seljak, mangano, hamann, wmap7}.

Another possible high redshift discrepancy, also sensitive to
relativistic degrees of freedom, involves primordial $^4$He
measurements compared with the predictions of Big Bang nucleosynthesis
(see, e.g., \cite{hamann, krauss}).  While standard BBN, assuming
a value of the baryon-photon ratio of $\eta=(6.19\pm0.15) \times
10^{-10}$ as measured by CMB data \cite{wmap7}, predicts a primordial
Helium mass fraction $Y_p=0.2487\pm0.0002$, current observational
measurements prefer a larger value of $Y_p=0.2561\pm 0.0108$
\cite{Aver:2010wq} and $Y_p=0.2565\pm 0.0010$(stat.)$\pm
0.0050$(syst.) \cite{it}.  Since the primordial $^4$He mass fraction
is largely determined by the neutron to proton ratio at the start of
BBN, $Y_p$ is sensitive to the value of the expansion rate and so
through the Friedmann equation to the overall energy density at
temperature $\sim 1\,MeV$, e.g.\ in relativistic particles. From
\cite{steigman} one has approximately $\Delta Y_p \simeq 0.013
(\neff-3)$ for $|\neff-3|\lesssim 1$.  Thus, an increase in the effective
number of relativistic degrees of freedom could remove the tension
between BBN and the measured $^4$He abundances.

These new relativistic degrees of freedom (RDOF) could be due, for
example, to a fourth (or fifth), sterile neutrino (a postulated 
neutrino species that does not participate in weak interactions).  
Such a hypothesis is worthwhile testing and may also
be compatible with recent neutrino oscillation results
(e.g.\ \cite{holanda,akhmedov,giunti}).  This origin would have no
direct relation to the question of cosmic acceleration and dark
energy.  

However, the signal could also arise from the class of early dark
energy models.  Scalar fields from dilatons in field theory and moduli
in string theory are generally predicted to possess scaling
properties, so that they would evolve as radiation in the radiation
dominated era and contribute a constant fraction of energy density.
Extending probes of dark energy to high redshift is an important
frontier and detection of its effects would provide an invaluable
guide to the physics behind cosmic acceleration.  Dark energy that is
significantly present not only in the late universe but also at early
times is called early dark energy (EDE; see, e.g.,
\cite{Doran:2006kp,Linder:2008nq}).  Furthermore, some theories,
typically involving higher dimensions, predict a ``dark radiation''
component in addition to a cosmological constant.

All three origins increase RDOF but have different evolutions in the energy density 
as radiation domination wanes, and hence will have the expansion history and $\neff$ 
differing as a function of time.  In this paper we consider the effects of contributions 
to $\neff$ from both neutrino and dark energy components, individually and together, 
and analyse the constraints imposed by current and future cosmological data.  
The main motivations are 1) to investigate how the current indication for RDOF would 
translate into a signal for dark energy at high redshift, and 
2) to examine how adopting a dark energy component that is not negligible at high 
redshift would impact the stability of future possible conclusions from Planck about RDOF. 

As a product of this analysis, we update the current constraints on EDE with 
recent data. Previous analysis have placed constraints on EDE using the available 
cosmological datasets and forecasting the discriminatory power of future 
CMB probes like Planck (see e.g.\ \cite{Calabrese:2010uf,dePutter:2010vy,Alam:2010tt, 
Hollenstein:2009ph,Xia:2009ys}). Here we revise these constraints by using updated 
datasets, by enlarging the parameter space through including shear viscosity in EDE 
perturbations, and by considering the possible degeneracies with sterile neutrinos.  
Section~\ref{sec:theory} describes in more detail the models used to give extra $\neff$.  
In Sec.~\ref{sec:analysis} we present the results of our analysis for the different cases, 
and Sec.~\ref{sec:discussion} discusses the conclusions about the ability to constrain and 
distinguish the various physical origins.

\section{Neutrinos, Early Dark Energy, and Dark Radiation} 
\label{sec:theory}

We consider three models that contribute to RDOF: sterile neutrinos, 
early dark energy, and a barotropic dark energy model that produces a 
dark radiation component in the early universe.
Accounting for each possible contribution, the RDOF translated into an 
effective number of neutrinos is :
\begin{equation}
\neff=\neffnu+\Delta\neff^{EDE}+\Delta\neff^B \ ,
\end{equation}
where $\neffnu$ is the number of neutrino species (including extra sterile
neutrinos), $\Delta \neff^{EDE}$ is the contribution coming from 
early dark energy, and $\Delta \neff^B$ is the contribution 
from barotropic dark energy.  
When the components do not behave completely relativistically, 
the effective numbers may be functions of redshift; for EDE the 
contribution is constant only well before matter-radiation equality, 
while barotropic dark energy behaves as a relativistic component at 
all times during and before recombination.  
In the following sections we describe 
our modelling for these three components and then their physical 
signatures in CMB power spectra, before proceeding to 
the ability of cosmological data to constrain and discriminate among them.

\subsection{Relativistic Neutrinos} 

For the purposes of exploring a deviation from the standard value of
$\neff=3.046$, we first assume thermal (so the factors in
Eq.~(\ref{eq:neffdef}) hold), massless sterile neutrinos which give a
time-independent contribution to $\neff$ according to
Eq.~(\ref{eq:neffdef}).  We indicate the total neutrino contribution
to $\neff$ by $\neffnu$, which is not necessarily equal to $\neff$ any
more.  Current cosmological data bound the mass of extra (thermal)
sterile neutrinos to be $m_{\nu,s}\lesssim0.5\,$eV for $\neff \ge 4$
at $95 \%$ c.l.~(\cite{Giusarma:2011ex,hamann}).  Such massive sterile
neutrinos may also be compatible with recent neutrino oscillation
results (e.g.\ \cite{holanda,akhmedov,giunti}), however considering
them as massive has negligible impact on the constraints on $\neff$
(again, see \cite{Giusarma:2011ex}) and we treat these neutrinos as
massless in what follows.  Neutrinos with $\sim$keV masses, sometimes
considered for sterile neutrinos, are non-relativistic at recombination
and contribute little to $\neff$ at that time.  Note that, while we
refer $\neffnu$ to sterile neutrinos in what follows, other
relativistic backgrounds (for example, gravitational waves) produce
identical effects on cosmology.  
See \cite{Galli:2008hm} for a decaying particle scenario.  
Any such model that lacks significant
contribution to late time energy density is for our purposes
equivalent to sterile neutrinos.

The usual case, e.g.~in \cite{act}, is to analyze the constraints on 
$\neffnu$ in the absence of an early dark energy density, in which 
case $\neff=\neffnu$.  We consider in this article $\neffnu$ in the 
presence of an early dark energy density.

\subsection{Early Dark Energy} 

Early dark energy is the name given to a dark component that in the 
recent universe acts to accelerate expansion, but which retains a 
non-negligible energy density at early times (e.g.\ around recombination, 
or earlier).  To keep an appreciable energy density at early times, the 
equation of state would not be negative, but at or near that of the 
background equation of state.  We adopt the commonly used form 
\cite{Doran:2006kp} 
\begin{eqnarray}
\od(a) &=&  \frac{\od^0 - \omegae \left(1- a^{-3 w_0}\right) }{\od^0 + \Omega_{m}^{0} a^{3w_0}} + \omegae \left (1- a^{-3 w_0}\right) \label{eq:edeom}\\
w(a) &=& -\frac{1}{3[1-\od(a)]} \frac{d\ln\od(a)}{d\ln a} + \frac{a_{eq}}{3(a + a_{eq})} 
\label{eq:edew} 
\end{eqnarray}
where $\od(a)$ is the fractional energy density and $w(a)$ the
equation of state of EDE.  The factor $\aeq/(a+\aeq)$ in
Eq.~(\ref{eq:edew}) comes from $\Omega_r(a)/[1-\od(a)]$ where $\Omega_r$
is the fractional radiation energy density (specifically excluding any
EDE) and $\aeq$ is the scale factor at matter-radiation equality.
Here $\od^0$ and $\Omega_{m}^{0}$ are the current dark energy and
matter density, respectively, and a flat Universe is assumed so
$\Omega_{m}^{0}+\od^0 = 1$.  The present equation of state
$w(a=1)=w_0$.  The energy density $\od(a)$ goes to a finite constant
$\ome$ in the past, in both the matter dominated and radiation dominated 
eras, indicating a scaling solution. 

The dark energy equation of state $w(a)$ follows three distinct
behaviours: $w\approx 1/3$ during radiation domination, $w\approx0$
during matter domination, and $w \approx w_0$ in recent epochs.  An
accurate fitting formula for the time variation of the EDE equation of
state during the recent universe is $w_a=5\ome$ \cite{Linder:2008nq}, 
where $w(a)=w_0+w_a(1-a)$ fits observational quantities at $z<3$ to
0.1\% accuracy.  We extend the model by modeling the EDE clustering
properties through the effective sound speed $c_s^2 = \delta p/ \delta
\rho$ and a viscosity parameter $c_{vis}^2$ that describes the
possible presence of anisotropic stresses (see
e.g.\ \cite{Calabrese:2010uf} and references therein).  In what
follows we assume these clustering parameters as constant and consider
two cases: $c_s^2=c_{vis}^2=1/3$, corresponding to a relativistic
origin, and $c_s^2=1$, $c_{vis}^2=0$ as expected in the case of a
quintessence scalar field.  For simplicity we consider $w_0=-1$ since the low
redshift data are consistent with a cosmological constant and viable
EDE models have little time variation there.

Regarding the contribution to $\neff$, the EDE scaling behavior
indicates the energy density will behave as a relativistic component
until the epoch of matter-radiation equality and so 
$\Delta\neff^{EDE}$ will be constant by Eq.~(\ref{eq:neffdef}).  
However, since $w$ then deviates from
$1/3$ toward 0 as the EDE later behaves more non-relativistically,
$\Delta \neff^{EDE}$ will grow.  Translating the EDE density
into an effective ``neutrino" number $\Delta \neff^{EDE}$ through 
Eq.~(\ref{eq:neffdef}) yields :
\be 
\Delta\neff^{EDE}(a) = \Big[\frac{7}{8} \Big( \frac{4}{11}
  \Big)^{4/3}\Big]^{-1} \frac{\rho_{de}(a)}{\rho_\gamma(a)} \ .  \ee

This is clearly a redshift-dependent quantity since as the EDE
equation of state begins to evolve differently from radiation the density 
ratio will vary with time.  In the limit $a\ll a_{eq}$, 
\be
\Delta \neff^{EDE}(a\ll a_{eq}) = 7.44 \frac{\ome}{1-\ome} \,, 
\label{eq:dneff} 
\ee 
since $\rho_{de}/\rho_\gamma=(\rho_{de}/\rho_{\rm rad})\,(\rho_{\rm rad}/ 
\rho_\gamma)$ and $\rho_{\rm rad}/\rho_\gamma=1.69$ for three neutrino 
species. 

\begin{figure}[h!]
\includegraphics[width=\columnwidth]{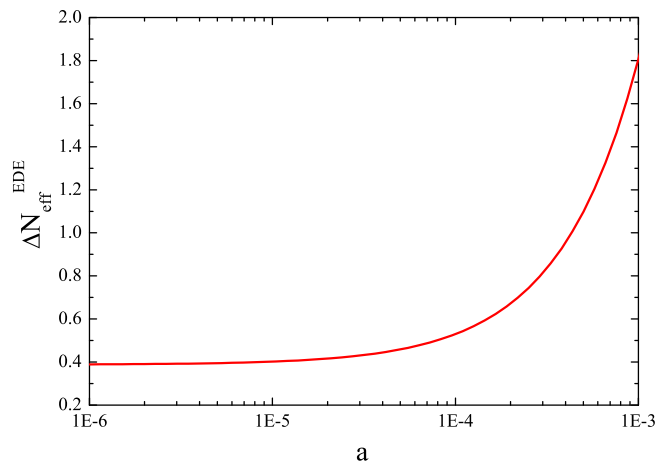}
\caption{Evolution of $\Delta \neff^{EDE}$ as a function of the scale
  factor $a$, for $\ome=0.05$ (the results scale nearly linearly for 
smaller values).  Note the strong time dependence near recombination. }
\label{delta_n}
\end{figure}

In Figure~\ref{delta_n} we plot $\Delta\neff^{EDE}(a)$, for
$\Omega_e=0.05$. As we can see, at early times EDE behaves like a RDOF
component with a constant value of $\Delta \neff^{EDE} \approx
0.39$. However this value increases at later times, when EDE starts to
mutate into a matter-scaling component, reaching $\Delta \neff^{EDE}
\approx 1.6$ at recombination.  This time dependence will be a crucial
element in discriminating between EDE and a sterile neutrino
contribution $\neffnu$.  Having $\Delta\neff$ in EDE models smaller at
BBN than at recombination helps ease the discrepancy between the lower
value expected (3.046) and that derived from CMB data.  Furthermore,
the larger value of $\Delta\neff$ at recombination means that the
constraints on EDE from CMB anisotropies will translate to tighter
bounds on $\Delta \neff^{EDE}$ at BBN, and hence on $Y_p$, than those
in the neutrino RDOF case.

\subsection{Barotropic Dark Energy} 

As a model with some characteristics of each of the previous two, we consider a case 
containing dark radiation, that is a component whose energy density evolves as 
$\rho\propto a^{-4}$ but does not interact electromagnetically.  Such terms arise in higher 
dimensional theories with multiple branes that induce a Weyl tensor 
contribution to the energy-momentum tensor \cite{Binetruy:1999hy}. 
Here we choose a more conventional model with interesting properties, arising from 
the barotropic class \cite{Scherrer:2005je}
where the pressure is an explicit function of the energy density.  
Barotropic models were shown to be highly predictive, reducing 
the cosmological constant fine tuning problem by rapidly evolving through an attractor 
mechanism toward $w=-1$ \cite{Linder:2008ya}.  Their equation of state, and hence energy 
density evolution, is wholly determined by their sound speed, through 
\be 
w'\equiv dw/d\ln a =-3(1+w)(c_s^2-w) \,. 
\ee 
This gives for any constant $c_s$ a time dependent equation of state 
\be 
w=[c_s^2 B a^{-3(1+c_s^2)} - 1] / [Ba^{-3(1+c_s^2)} + 1] \ , 
\ee 
where $B=(1+w_0)/(c_s^2-w_0)$.  

Being interested in relativistic degrees of freedom, we choose $c_s^2=1/3$ (and 
indeed $c_s^2>1/3$ would violate early radiation domination).  This leads to 
a surprisingly simple solution: 
\be
\rho_{\rm baro}(a)=\rho_\infty+C \rho_{r,0} a^{-4} \ , 
\ee
where $\rho_\infty=(3H_0^2/8\pi G)(1-\om-C\Omega_{r,0})$ and $C=\ome^B/(1-\ome^B)$.  
This acts like radiation at early times, with a constant fractional energy density $\ome^B$ 
during the radiation dominated era.  
At late times it has a constant absolute energy density $\rho_\infty$.  It basically looks 
like the sum of a cosmological constant and dark radiation, despite having no explicit 
cosmological constant.  
As expected, at early times $w=1/3$ and at late times 
$w$ rapidly approaches $-1$.   We take $w_0=-0.99$ (since $w=-1$ is only reached 
asymptotically), and $c_{vis}^2=1/3$ to match the other cases. 

To clearly state the main practical differences of our three models: 
extra neutrino species give a constant contribution to 
$\neff$ and negligible contribution to late time energy density as well 
as no acceleration; standard early dark energy gives a time varying 
contribution to $\neff$ as well as late time energy density and 
acceleration; barotropic dark energy has the third interesting combination 
of properties, giving a constant contribution to $\neff$ but also late 
time energy density and acceleration.  The interplay between these 
properties will allow each model to impact the observations in a 
distinctive manner. 

In addition to approaching this model microphysically, through the class 
of barotropic models, one can obtain an equivalent result within k-essence 
\cite{Scherrer:2004au} using the quadratic Lagrangian $\mathcal{L}=X_0+cX^2$, 
where $X$ is the kinetic energy and $c$, $X_0$ are constants.  

Because for this model the relativistic scaling occurs so quickly (by
$z>5$), the $\Delta \neff^B$ contribution in this case will be
constant at and before recombination, like the neutrino model with
$\Delta\neff^B=7.44\,\ome^B/(1-\ome^B)$ (see Eq.~\ref{eq:dneff}).
However it has the late time change in equation of state that will
affect large scale aspects of the CMB, and other cosmological probes,
like the EDE model.  Thus we expect the results to have aspects of each of 
the other two cases.

\subsection{Effects of the new components}

\begin{figure}[h!]
\includegraphics[width=\columnwidth]{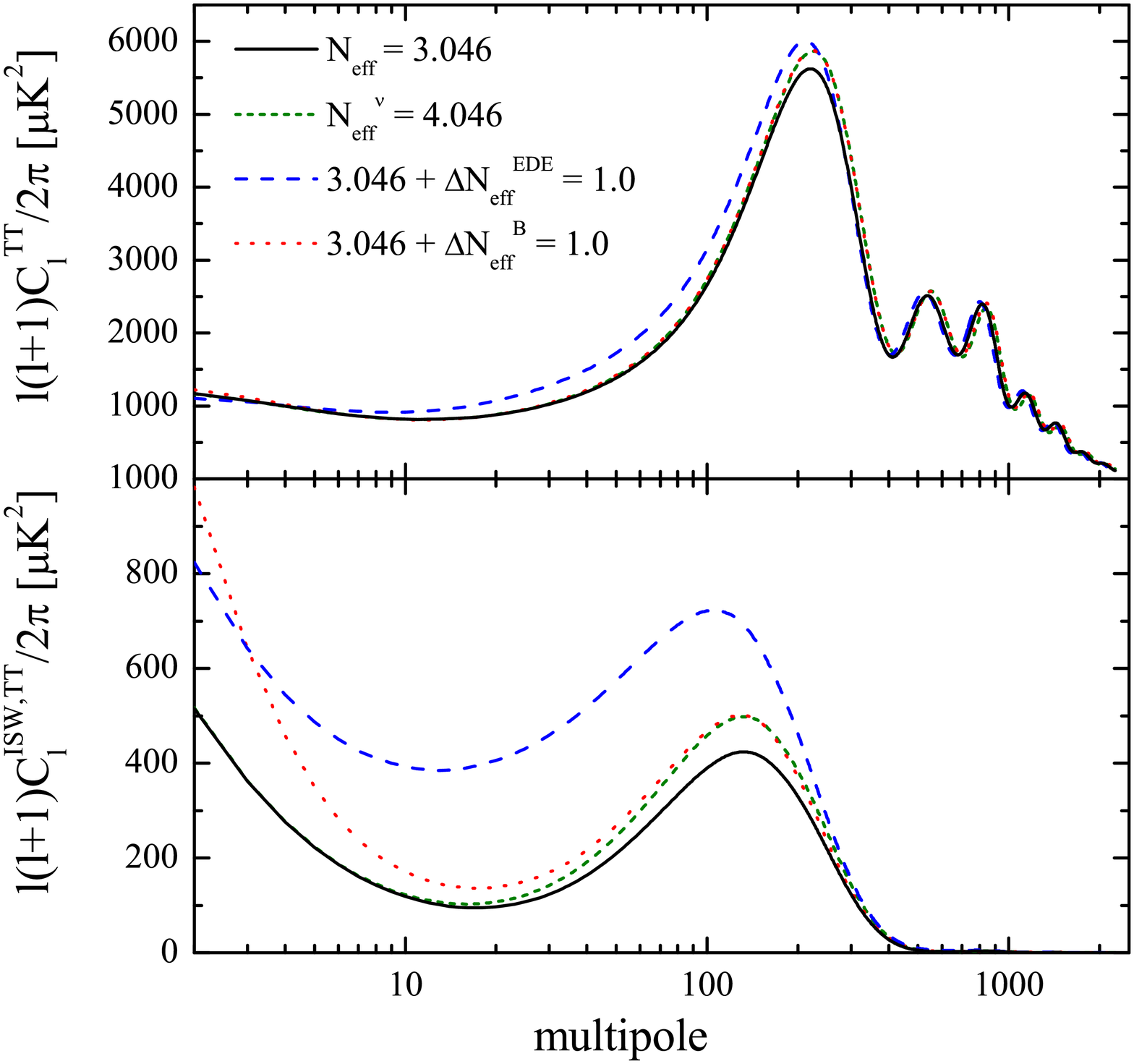}
\caption{CMB temperature (top panel) and ISW contribution alone
  (bottom panel) angular power spectra dependence from sterile
  neutrinos, early dark energy and barotropic dark energy.  All the
  models have been chosen to produce one extra relativistic degree of
  freedom at the epoch of BBN, except for the solid curve showing
  the standard case.} 
\label{spettri}
\end{figure}

Since our main observational datasets are CMB anisotropies, it is
useful to see the effects of sterile neutrinos, EDE and barotropic
dark energy on the CMB anisotropy angular spectrum.  In Figure
\ref{spettri} (top panel) we show the CMB temperature angular spectra
for these $3$ components assuming that they contribute at the level of
one extra degree of freedom at BBN.  As we can see, while sterile
neutrinos and barotropic dark energy produce nearly identical angular
spectra, EDE predicts a significantly different spectrum. This is
clearly shown in the bottom panel of the same figure where we plot the
isolated Integrated Sachs Wolfe (ISW) contribution for each case.  As
we can see, the main difference between a sterile neutrino and EDE
comes from the ISW effect. This is mainly due to the time-dependence
of the equation of state in the EDE component that tracks the dominant
component at all epochs and increases the ISW signal on all angular
scales.  At the same time, we see that barotropic dark energy differs
from a sterile neutrino in the increase in the ISW at large angular
scales, due to the variation in the equation of state at small
redshift in the barotropic component.

\section{Analysis and Results} 
\label{sec:analysis} 

We perform a COSMOMC \cite{camb} analysis combining the following CMB
datasets: WMAP7 \cite{wmap7}, ACBAR \cite{acbar}, QUAD \cite{quad}
(collectively referred to as ``All''), and ACT \cite{act}.  We analyze
datasets using out to $l_{\rm max}=2500$.  We also include 
information on dark matter clustering from the galaxy power spectrum
extracted from the SDSS-DR7 luminous red galaxy sample
\cite{red}. Finally, we impose a prior on the Hubble parameter based
on the Hubble Space Telescope observations \cite{hst}.

The analysis method we adopt is based on the publicly
available Markov Chain Monte Carlo package \texttt{cosmomc}
\cite{Lewis:2002ah} with a convergence diagnostic done through the Gelman and
Rubin statistic.  We sample the following nine-dimensional set of
cosmological parameters, adopting flat priors on them: the baryon and cold
dark matter densities $\omega_{\rm b}$ and $\omega_{\rm c}$, the Hubble
constant $H_0$, the scalar spectral index $n_S$, the overall normalization of
the spectrum $A$ at $k=0.05\Mpc^{-1}$, the SZ amplitude $A_{SZ}$, 
the optical depth to reionization, $\tau$, the effective number of relativistic
neutrinos $\neffnu$, 
and finally the early density $\ome$, for either the case of EDE or the
barotropic model ($\ome^B$).
For the ACT dataset we also consider two extra parameters accounting for 
the Poisson and clustering point sources foregrounds components.
We consider purely adiabatic initial conditions and we impose spatial flatness. 

To study the impact of EDE perturbations, we consider 
two cases: ``quintessence" ($\cs^{2}=1, \cvis^2=0$) and 
 ``relativistic"  ($\cs^{2}=1/3=\cvis^2$) EDE scenarios.
For the barotropic dark energy model we assume $\cs^{2}=1/3=\cvis^2$.

\begin{table*}[htb!]
\begin{tabular}{|c|c|c|c|c|}
\hline
& \multicolumn{4}{|c|}{} \\
 & \multicolumn{4}{|c|}{ \textbf{All+ACT}}\\
\hline
& \multicolumn{2}{|c|}{} & \multicolumn{2}{|c|}{} \\
Model: & \multicolumn{2}{|c|}{ \textbf{$\neffnu = 3.046$ }} & \multicolumn{2}{|c|}{ \textbf{ $\neffnu$ varying}} \\
\hline
& & & & \\
& $\cs^{2}=\cvis^2=1/3$ & $\cs^{2}=1, \cvis^2=0$ & $\cs^{2}=\cvis^2=1/3$ & $\cs^{2}=1, \cvis^2=0$ \\
\hline
& & & & \\
Parameter & & & & \\
\hline
& & & &\\
$\Omega_b h^2$ & $0.02218 \pm 0.00044$ & $0.02232 \pm 0.00044$ & $0.02238 \pm 0.00047$ & $0.02259 \pm 0.00048$\\
$\Omega_c h^2$ & $0.1178 \pm 0.0039$ & $0.1163 \pm 0.0038$ & $0.138 \pm 0.012$ & $0.139  \pm 0.011$\\
$H_0$ & $68.2 \pm 1.7$ &  $67.8 \pm 1.6$ & $72.5 \pm 2.8$ & $72.4 \pm 2.7$ \\
$n_s$ & $0.971 \pm 0.013$ & $0.964 \pm 0.011$ & $0.988 \pm 0.015$ & $0.986 \pm 0.015$\\
$t_0/\Gyr$ & $13.71 \pm 0.30$ & $13.83 \pm 0.29$ & $12.91 \pm 0.48$ & $12.94 \pm 0.48$\\
$\neffnu$ & $3.046$ & $3.046$ & $4.37 \pm 0.76$ & $4.49 \pm 0.72$\\
$\Omega_e$ & $< 0.043$ & $< 0.024$ & $< 0.039 $ & $<0.020$\\
$\Delta \neff^{EDE}(a_{\rm BBN}) $ & $<0.34$ & $<0.18$ & $< 0.32$ & $<0.18$\\
$Y_p$ & $0.2504 \pm 0.0013$ & $0.2495 \pm 0.0008$ & $0.2661 \pm 0.0078$ & $0.2667 \pm 0.0080$ \\
& & & & \\
\hline
\end{tabular}
\caption{Best-fit values and $68 \%$ confidence errors on cosmological
  parameters using the current cosmological data. For
  $\omegae$ and $\Delta \neff^{EDE}(a_{\rm BBN})$, EDE density and the contribution to
  the RDOF from EDE at the BBN epoch respectively, the upper bounds at $95\%$
  c.l. are reported. See text for other details.}
\label{mcmc_results}
\end{table*}

\subsection{Constraints on EDE with $\neffnu=3.046$ fixed}

We first perform an analysis of current data fixing the effective
number of relativistic neutrinos to the standard value of
$\neffnu=3.046$ and varying the amount of early dark energy,
parametrized as $\Omega_e$.  The EDE affects the CMB angular power
spectrum at all multipoles as shown in Fig.~\ref{spettri}. 
We convert the EDE into an equivalent additional relativistic species $\Delta
\neff^{EDE}$ and quote this parameter at the BBN epoch. 
We also recognize the fact that the changed
expansion rate during the BBN, due to the presence of EDE, also
alters the primordial Helium mass fraction $Y_p$, and show the
effect of $\Omega_e$ on $Y_p$ below. 

 The MCMC results on the cosmological parameters are reported in the
first two columns of Table \ref{mcmc_results}. 

As we can see, the cosmological data we consider do not provide
evidence for an EDE component and significantly improve the bound
\cite{dePutter:2010vy}, yielding a $95 \%$ c.l.\ upper
limit of $\Omega_e < 0.043$ in case of a ``relativistic" EDE with
$\cs^{2}=\cvis^2=1/3$, and a bound of $\Omega_e < 0.024$ in case of a
``quintessence" EDE with $\cs^{2}=1$, $\cvis^2=0$.  The assumptions of
$\cs$ and $\cvis$ strongly affect the bounds on $\Omega_e$. The
``quintessence" scenario leaves a stronger signal on the CMB
anisotropies (see e.g.~\cite{Calabrese:2010uf}) and is therefore
better constrained.  This arises from the greater decay of potentials 
at recombination in contrast to the low sound speed case where the 
dark energy perturbations help sustain the potentials. 

In order to investigate the impact of the recent ACT dataset, which
samples very small angular scales, giving a long lever arm, on the
final result we also perform an analysis excluding it in the case of
``relativistic" EDE. 
Without ACT we get a $\sim20\%$ weaker bound, $\omegae < 0.051$ at
$95 \%$ c.l..

While the EDE component is not preferred, it is also not excluded from
current data. It is therefore interesting to investigate if the EDE
component compatible with cosmological data is also compatible with
BBN data. For this reason we translated the bounds on $\Omega_e$ into
the corresponding bound on $\Delta \neff^{EDE}$ expected at time of
onset of BBN, using Eq.~(\ref{eq:dneff}), and computed the expected
abundance $Y_p$ in primordial $^4$He by making use of the public
available PArthENoPE BBN code (see \cite{iocco}). In other
  words, the constraints on early dark energy during the BBN
  correspond to limits on the expansion rate of the universe at this
  epoch, which translate into the corresponding limits on the excess
  of primordial mass fraction of Helium.

\begin{figure}[h!]
\includegraphics[width=\columnwidth]{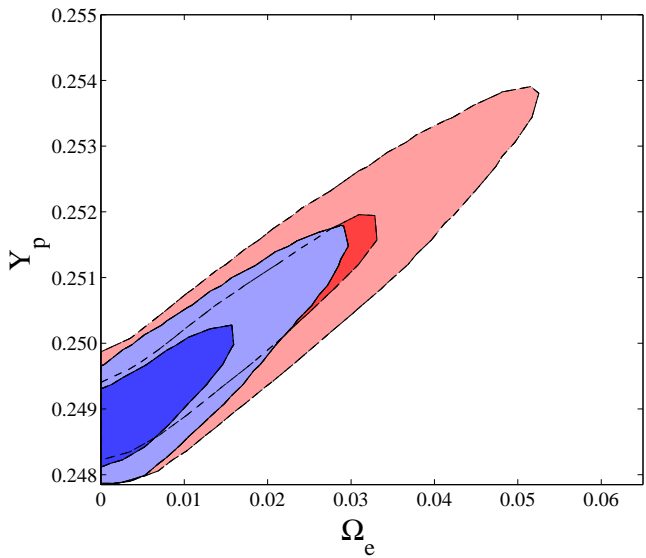}
\caption{68\% and 95\% c.l.\ contours in the $Y_p$-$\omegae$ plane for
  the standard EDE model. The red dashed contours show the
  $\cs^{2}=\cvis^2=1/3$ model, while the blue solid contours show the
  $\cs^{2}=1, \cvis^2=0$ model.  Since the early dark energy enhances
  the expansion rate during the BBN, it allows for a higher primordial
  Helium mass fraction according to $\Delta Y_p \simeq 0.013
  (\neff-3)$ \cite{steigman}.  }
\label{yp_oe}
\end{figure}

Table~\ref{mcmc_results} shows the ``relativistic'' case $95 \%$ upper
limit $\Omega_e<0.043$ translates to a $95 \%$ constraint of $Y_p =
0.2504\pm0.0026$ (with $\Delta \neff^{EDE}< 0.34$), while the
``quintessence'' case $95 \%$ upper limit $\Omega_e<0.024$ translates
to a $95 \%$ constraint of
$Y_p = 0.2495 \pm0.0016$ (with $\Delta \neff^{EDE} < 0.18$). These
values should be compared with the theoretical value of
$Y_p=0.2487\pm0.0002$ obtained assuming standard BBN and $\Omega_e
=0$.  EDE is therefore clearly shifting the BBN predictions on $Y_p$
towards larger values with weaker constraints.  The weaker constraints
indicate a degeneracy between $\ome$ and $Y_p$, as we discuss more in
the next section, that we also show in Figure \ref{yp_oe} where we
plot the $68 \%$ and $95 \%$ constraints on the $Y_p$-$\Omega_e$ plane
in the case of ``relativistic" or a ``quintessence" EDE.

As stated in the introduction, current experimental measurements seems
to prefer a larger value for the primordial Helium with $Y_p=0.2561\pm
0.0108$ (see \cite{Aver:2010wq}) or $Y_p=0.2565\pm 0.0010$ (stat.)
$\pm 0.0050$ (syst.) from \cite{it}.  These results are off by $\sim
1.5\sigma$ from the expectations of standard BBN but introducing EDE
acts to alleviate this tension. Given the possibility of systematics
in measuring the primordial nuclear abundances, however, it is
premature to derive any conclusion.

\subsection{Constraints on EDE and $\neffnu$}

As a second step, we include into the analysis the possibility of
extra sterile neutrinos, parametrizing it with the effective neutrino
number $\neffnu$.  As we can see from Table~\ref{mcmc_results} (last
two columns), the constraints on $\ome$ are practically unaffected by
the inclusion of extra RDOF and vice versa.  From our analysis we found
that sterile neutrinos are preferred with $\neffnu = 4.37 \pm 0.76$ at
$68 \%$ c.l..  This constraint should be compared with the bound from
the analysis of \cite{act} of $\neffnu=4.6\pm0.8$ at $68 \%$ c.l.,
obtained with similar datasets but without EDE, indicating that the
effect of EDE on the constraint is small.  The low covariance between
the number of sterile neutrinos and EDE density comes from the
property that while at BBN they both act as RDOF, by recombination
the EDE behaves more like matter and so can be constrained separately
from the neutrino contribution.

\begin{figure}[h!]
\includegraphics[width=\columnwidth]{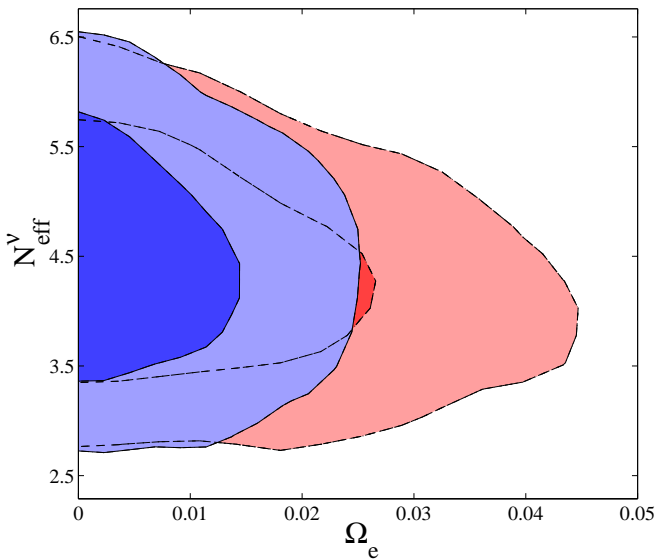}
\caption{68\% and 95\% c.l.\ contours in the $\neffnu$-$\omegae$
  plane for the standard EDE model plus neutrinos).  The red
  dashed contours refer to $\cs^{2}=\cvis^2=1/3$ case, while the blue
  solid contours refer to the $\cs^{2}=1, \cvis^2=0$ case.  }
\label{nnu_oe}
\end{figure}

This can also be seen in Figure~\ref{nnu_oe} where we show the
likelihood contour plots in the $\neffnu$-$\omegae$ plane for the
cases of ``relativistic" and ``quintessence" EDE. There is no strong
degeneracy between $\Omega_e$ and $\neffnu$.  This, together with the small
value of $\Delta \neff^{EDE}$ allowed, 
indicates that the current hints for the existence of the extra RDOF
cannot be completely explained by a conventional EDE model.  In the
next subsection we will see that the barotropic class of early dark
energy has more success.

As we can see from Table~\ref{mcmc_results}, including the possibility of 
extra neutrino 
contributions to $\neff$ greatly enlarges the CMB bounds on primordial
Helium, with $Y_p = 0.2661\pm0.0078$ in case of ``relativistic" EDE to
$Y_p = 0.2667\pm0.0080$ in case of ``quintessence" EDE. This 
stronger influence of neutrino RDOF than EDE has the consequence 
that in this situation the impact of EDE on the 
$Y_p$ abundance is small. As seen from the results in
Table~\ref{mcmc_results}, the $\Delta \neff^{EDE}$ from EDE at BBN is always better
constrained from CMB data than the $\neffnu$ expected from a
sterile neutrino.  If future measurements of primordial $^4$He clearly
point towards value of $Y_p\sim 0.26$, it will not be possible to
explain this result with a conventional EDE contribution.

Finally, we note that including the possibility of
$\neffnu > 3$ also changes the constraints on $n_s$ making it
more compatible with a Harrison-Zeldovich, $n_s=1$, primordial
spectrum (cf.~\cite{Galli:2008hm, DeBernardis:2008tk}).  
The best fit value of the inflationary 
tilt $n_s-1$ is reduced by almost a factor of 3, which would have 
a substantial impact in the reconstruction of the
inflationary potential.

\subsection{Results on Barotropic Dark Energy}

\begin{figure}[h!]
\includegraphics[width=\columnwidth]{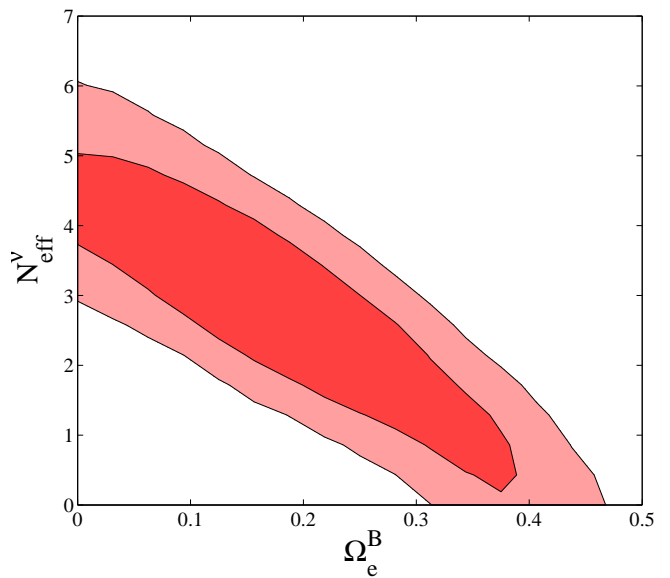}
\caption{68\% and 95\% c.l.\ contours in the $\neffnu$-$\omegae^B$
  plane for the barotropic dark energy model.}
\label{neff_oe_baro}
\end{figure}

The barotropic model contributes both early dark energy
density and a constant (rather than diluted as in Fig.~\ref{delta_n}) 
early time RDOF.  This will have interesting implications.  
As for the conventional EDE case, we add the early
density, here $\ome^B$, to the MCMC analysis to estimate constraints
on cosmological parameters.  We also allow $\neffnu$ to vary as in the
previous section.  For simplicity, we otherwise fix $w_0 = -0.99$ and
$\cs^2 = \cvis^2 = 1/3$.  The results are reported in
Table~\ref{mcmc_baro}, and in Figure~\ref{neff_oe_baro} we show the
degeneracy between $\neffnu$ and $\omegae^B$ parameters.

The barotropic model strongly alters the constraints on $\neffnu$ and
a non-negligible presence of the dark radiation part of the barotropic
dark energy at recombination could not only bring the constraints on
$\neffnu$ back in agreement with the standard value of $\neffnu=3.046$
but even erase the current claim for a neutrino background from CMB
data. A ``neutrinoless'' model with $\neffnu=0$ and $\omegae^B=0.4$,
albeit extreme, is allowed by the cosmological data we consider here.

\begin{table}[htb!]
\begin{center}
\begin{tabular}{|c|c|}
\hline
& \\
Parameter & All + ACT \\
\hline
& \\
$\Omega_b h^2$ & $0.02209 \pm 0.00055 $ \\
$\Omega_c h^2$ & $0.135 \pm 0.012$ \\
$H_0$ & $71.1 \pm 2.8$ \\
$n_s$ & $0.986 \pm 0.015$ \\
$t_0/\Gyr$ & $13.18 \pm 0.51$ \\
$\neffnu$ & $ < 5.1$ \\
$\Omega_e^B$ &  $< 0.37$ \\
$\Delta \neff^{B} $ & $<2.8$\\
$Y_p$ & $0.2649 \pm 0.0084$ \\
& \\
\hline
\end{tabular}
\caption{Best-fit values, together with $68 \%$ confidence errors, on
  cosmological parameters for the barotropic model using current data. 
  For the $\neff$, $\omegae^B$ and $\Delta \neff^B$ parameters
  upper bound at $95\%$ c.l.\ are reported.} 
\label{mcmc_baro}
\end{center}
\end{table}

As in the case for $\neffnu > 3$, when a barotropic dark energy model 
is considered (even without extra neutrinos) a high value of $Y_p$ is 
consistent and the constraints on $n_s$ are moved toward a 
Harrison-Zeldovich primordial spectrum.

\subsection{Forecasts for the Planck Satellite Mission}

Looking to the future, we investigate the constraints on EDE and RDOF 
in the case of the data as expected from the Planck satellite.
To evaluate the future constraints achievable from this satellite,
we consider an experimental configuration with 
three frequency channels with the 
specifications as listed in Table~\ref{exp} (see \cite{planck}).

\begin{table}[!htb]
\begin{center}
\begin{tabular}{ccccc}
Experiment & Channel[GHz] & FWHM & $\sigma_T [\mu K]$ &  $\sigma_P [\mu K]$\\
\hline
Planck & 143 & 7.1'& 6.0 & 11.4\\
$f_{\rm sky}=0.85$       & 100 & 10.0'& 6.8 & 10.9\\
       & 70 & 14.0'& 12.8 & 18.3\\ 
\end{tabular}
\caption{Planck experimental specifications.} 
\label{exp}
\end{center}
\end{table}

For each frequency channel we consider a detector noise of
$(\theta\sigma)^2$ where $\theta$ is the FWHM of the beam assuming a
Gaussian profile and $\sigma$ is the sensitivity.  We therefore take 
a noise spectrum given by
\begin{equation}
N_\ell^X = (\theta\sigma_X)^2\,e^{l(l+1)/l_b^2} \, ,
\end{equation}
where $l_b \equiv \sqrt{8\ln2}/\theta$ and the label $X$ refers to either temperature
or polarization, $X=T, P$. \\

We perform the standard Fisher matrix analysis evaluating (see e.g.\
\cite{fishcmb}):
\begin{equation}
F_{ij}\equiv \Bigl\langle -\frac{\partial^2 
\ln \mathcal{L}}{\partial p_i \partial p_j}\Bigr\rangle _{p_0} \ ,
\end{equation} 
where $\mathcal{L}({\rm data}|{\bf{p}})$ is the likelihood function of
a set of parameters ${\bf p}$ given some data, and the partial
derivatives and the averaging are evaluated using the fiducial values
${\bf p_{0}}$ of the parameters.  The Cram\'er-Rao inequality implies
that $(F^{-1})_{ii}$ is the smallest variance in the parameter $p_i$,
so we can generally think of $F^{-1}$ as the best possible covariance
matrix for estimates of the vector ${\bf p}$. The one sigma error forecasted 
for each parameter is then given by $\sigma_{p_{i}}
= \sqrt{(F^{-1})_{i i}}$.

We consider a set of $10$ cosmological parameters with the following
fiducial values: the physical baryonic and cold dark matter densities
relative to critical $\Omega_b h^2=0.02258$ and $\Omega_c h^2=0.1109$,
the optical depth to reionization $\tau=0.088$, the Hubble parameter
$H_0=71\, {\rm km/s/\Mpc}$, the current dark energy equation of state
$w_0=-0.95$, the early dark energy density relative to critical
$\omegae=0.03$, the spectral index $n_s=0.963$, and the number of
relativistic neutrinos $\neff=3.046$.  For the last two
  parameters, the effective and viscous sound speeds $\cs^{2}$ and
  $\cvis^2$, we choose alternate fiducial values of $(1/3, 1/3)$ (the
  ``relativistic'' model) or $(1, 0)$ (the ``quintessence'' model).

\begin{enumerate}
\item {\textbf{EDE Forecasts}} \\ In Table~\ref{fisher} we report the
  uncertainties obtained on the cosmological parameters. The
  degeneracy between $\omegae$ and $\neffnu$ is shown in
  Figure~\ref{neff_oe} for the two analysed cases.  As seen in the
  Figure and in the Table, the future data from Planck will provide
  strong constraints on the RDOF: $\sigma(\neff)=0.11$, with little
  impact from the EDE density. If EDE with $\ome=0.03$ is present, it
  will be detected at high significance, since $\sigma(\ome)\approx
  0.005$.  The radiation and quintessence configurations of EDE can
  also be distinguished.

\begin{table}[!htb]
\begin{tabular}{|c|c|c|c|}
\hline
\multicolumn{2}{|l|}{} & \multicolumn{2}{|c|}{} \\
\multicolumn{2}{|l|}{} & \multicolumn{2}{|c|}{ \textbf{Planck $1$-$\sigma$ uncertainty}}\\
\hline
\multicolumn{2}{|l|}{} & & \\
\multicolumn{2}{|l|}{Model:} & $\cs^{2}=\cvis^2=1/3$ & $\cs^{2}=1, \cvis^2=0$ \\
\hline
& & & \\
Parameter & Fiducial & & \\
\hline
& & &\\
$\Omega_b h^2$ & $0.02258$ &$0.00016$ & $0.00014$\\
$\Omega_c h^2$ & $0.1109$ & $0.0018$ & $0.0017$ \\
$\tau$ & $0.0880$ & $0.0020$ & $0.0022$ \\
$H_0$ & $71.0$ & $8.5$ & $8.8$ \\
$n_s$ & $0.9630$ & $0.0046$ & $0.0044$ \\
$\neffnu$ & $3.046$ & $0.11$ & $0.11$\\
$w_0$ & $-0.95$ & $0.24$ & $0.24$ \\
$\Omega_e$ & $0.030$  & $0.005$ & $0.004$ \\
$\cs^{2}$ & $0.33$ & $0.047$ & $-$ \\
$\cvis^2$ & $0.33$ & $0.13$ & $-$ \\
$\cs^{2}$ & $1.00$ & $-$ & $0.34$ \\
$\cvis^2$ & $0$ & $-$ & $0.11$ \\
\hline
\end{tabular}
\caption{Fiducial errors and forecasted 1-$\sigma$ errors expected
  from the Planck satellite in the EDE scenario.}
\label{fisher}
\end{table}

\begin{figure}[h!]
\includegraphics[width=\columnwidth]{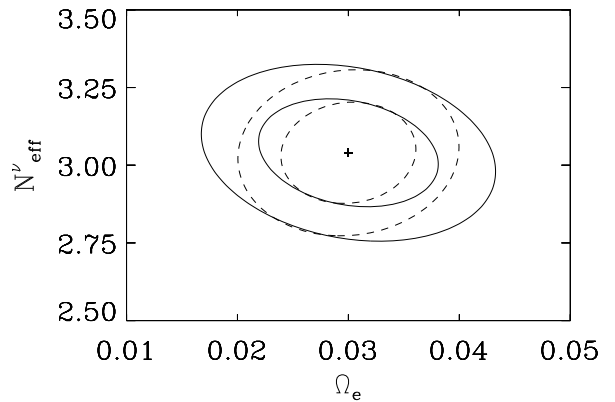}
\caption{68\% and 95\% c.l.\ constraints in the $\neffnu$-$\omegae$
  plane.  Solid lines show the ``relativistic'' case
  $\cs^{2}=\cvis^2=1/3$, while dashed lines show the ``quintessence''
  case $\cs^{2}=1, \cvis^2=0$.  The fiducial values are given by the
  ``+'' symbol.}
\label{neff_oe}
\end{figure}

\item{\textbf{Barotropic DE Forecasts }}\\ 
Similarly to the previous analysis, we forecasted the errors on
cosmological parameters with data expected from Planck in a barotropic
dark energy scenario. We report in Table~\ref{fisher_baro} the
$1$-$\sigma$ errors, and in Figure~\ref{neff_oe_fisher_baro} we show the
degeneracy between $\neffnu$ and $\omegae^B$.

\begin{table}[!htb]
\begin{tabular}{|c|c|c|}
\hline
\multicolumn{2}{|l|}{} & \\
\multicolumn{2}{|l|}{} & \textbf{Planck $1-\sigma$ uncertainty}\\
\hline
\multicolumn{2}{|l|}{} & \\
\multicolumn{2}{|l|}{Model:} & $\cs^{2}=\cvis^2=1/3$ \\
\hline
& & \\
Parameter & Fiducial &\\
\hline
& & \\
$\Omega_b h^2$ & $0.02258$ & $0.00013$ \\
$\Omega_c h^2$ & $0.1109$ & $0.0019$\\
$\tau$ & $0.0880$ & $0.0022$\\
$H_0$ & $71.00$ & $0.88$ \\
$n_s$ & $0.9630$ & $0.0041$ \\
$\neffnu$ & $3.046$ & $0.17$ \\
$w_0$ & $-0.95$ & $0.041$\\
$\Omega_e^B$ & $0.030$  & $0.015$ \\
$\cs^{2}$ & $0.33$ & $0.045$ \\
$\cvis^2$ & $0.33$ & $0.17$ \\

\hline
\end{tabular}
\caption{Same as Table~\ref{fisher}, but for the barotropic dark
  energy scenario.}
\label{fisher_baro}
\end{table}

\begin{figure}[h!]
\includegraphics[width=\columnwidth]{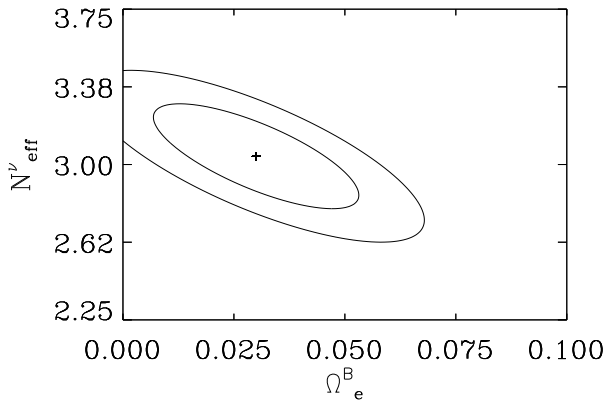}
\caption{2-D contour plots at 68\% and 95\% c.l. in the plane $\neffnu$-$\omegae^B$ 
for $\cs^{2}=\cvis^2=1/3$. The fiducial values are reported with the ``+'' symbol. }
\label{neff_oe_fisher_baro}
\end{figure}

For a fiducial early density of $\ome^B=0.03$, the barotropic model 
cannot be readily distinguished from the standard EDE model.  However, 
if $\ome^B$ is much larger, then distinction will be possible, 
with the associated implications for RDOF, $Y_p$, and $n_s$.
\end{enumerate}

\section{Conclusions}
\label{sec:discussion}

Data analyses of recent cosmological data have reported an interesting
indication for the presence of an extra background of relativistic
particles.  In this paper we have investigated the stability of this
result by considering the influence of a possible early dark energy
component.  We have found that current data do not provide evidence
for an EDE component, updating and strengthening previous constraints
on EDE, although there is still room for an interesting contribution.
In particular, we found the following $95 \%$ c.l.\ upper limits:
$\Omega_e<0.043$ for ``relativistic" EDE ($\cs^{2}=\cvis^2=1/3$) and
$\Omega_e<0.024$ for ``quintessence" EDE ($\cs^{2}=1$, $\cvis^2=0$).
These bounds translate into an extra relativistic background at BBN of
$\Delta \neff^{EDE} < 0.34$ and $\Delta \neff^{EDE} < 0.18$ at $95 \%$
c.l., respectively.

The EDE models are therefore not able to change the amount of
primordial $^4$He produced in BBN by more than $\Delta Y_P^{EDE} =
0.0044$, and do not help much in explaining why recent 
measurements of abundances of primordial Helium show values larger
than those expected from standard BBN.  The systematics in those 
measurements are however still too large to conclude that 
there is conflict between the measured and predicted $^4$He 
abundance. 

When {\it both\/} an EDE and extra sterile neutrinos are considered in
the analysis, there is very little degeneracy between them and the
constraints are virtually unaffected. The indication for extra
neutrinos in current data at about two standard deviations is
unchanged even after allowing a EDE component.  
We found $\neffnu=4.37\pm0.75$ at 68\% c.l.\ for
relativistic EDE and $\neffnu=4.49\pm0.72$ for quintessence EDE when
CMB, SDSS-DR7 and HST data are combined.  The bounds on $\Omega_e$ are
practically unchanged.  The key point is that EDE starts to behave
differently from a relativistic component after radiation-matter
equality, before the epoch of recombination. CMB data can therefore
provide crucial information in discriminating between $\neffnu$ and
early dark energy while for BBN these two components are virtually
indistinguishable.

However, when a barotropic dark energy model is considered, we have
found that the constraints on $\neffnu$ can be strongly altered,
bringing the standard value of $\neffnu=3.046$ back into perfect
agreement with observations.  In fact, even a ``neutrinoless" model
with $\neffnu=0$ and $\omegae^B=0.4$ is allowed given current
observations. While that is extreme, the model dependency clearly indicates
the caveats of considering $\neff >3.046$ as an indication for an
extra sterile neutrino or claiming any detection for a neutrino
background.  Interestingly, barotropic dark energy also shifts $Y_p$
to higher values, and reduces $|n_s-1|$, with implications for
models of inflation. 

Finally, we have shown that for the Planck experiment alone, again no
substantial degeneracy is expected between $\Omega_e$ and $\neffnu$,
with an expected accuracy of $\sigma(\neffnu)=0.11$ and
$\sigma(\Omega_e)=0.005$.  However, in a barotropic dark energy
scenario the degeneracy is present and Planck will offer weaker bounds
with an estimated $\sigma(\neffnu)=0.17$ and
$\sigma(\Omega_e^B)=0.015$. Both are still a great improvement over
present data.  Planck will therefore shed light on various scenarios
for early dark energy and relativistic degrees of freedom, exploring
if new physics exists in the neutrino or dark energy sectors,
a possible shift in the inflationary tilt $n_s-1$, and
the consistency of the primordial Helium abundance $Y_p$.

\acknowledgments

EC and AM are supported by PRIN-INAF grant, ``Astronomy probes
fundamental physics''.  DH is supported by the DOE OJI grant under
contract DE-FG02-95ER40899, NSF under contract AST-0807564, and NASA
under contract NNX09AC89G.  EL has been supported in part by the World
Class University grant R32-2009-000-10130-0 through the National
Research Foundation, Ministry of Education, Science and Technology of
Korea and also by the Director, Office of Science, Office of High
Energy Physics, of the U.S.\ Department of Energy under Contract
No.\ DE-AC02-05CH11231; he thanks Nordita for hospitality during part
of this work, and Claudia de Rham for helpful discussions.



\begin{thebibliography}{}
\bibitem{wmap7}
  E.~Komatsu {\it et al.},
  arXiv:1001.4538 [astro-ph.CO].

\bibitem{acbar}
  C.~L.~Reichardt {\it et al.},
  Astrophys.\ J.\  {\bf 694} (2009) 1200
  [arXiv:0801.1491 [astro-ph]].

\bibitem{quad}
  S.~Gupta {\it et al.}  [QUaD collaboration],
  arXiv:0909.1621 [astro-ph.CO].

\bibitem{act}
  J.~Dunkley {\it et al.},
  arXiv:1009.0866 [astro-ph.CO].

\bibitem{hamann}
  J.~Hamann, S.~Hannestad, G.~G.~Raffelt, I.~Tamborra and Y.~Y.~Y.~Wong,
  Phys.\ Rev.\ Lett.\  {\bf 105} (2010) 181301
  [arXiv:1006.5276 [hep-ph]].


\bibitem{mangano}
  G.~Mangano, A.~Melchiorri, O.~Mena, G.~Miele and A.~Slosar,
  JCAP {\bf 0703}, 006 (2007)
  [arXiv:astro-ph/0612150].

\bibitem{seljak}
  U.~Seljak, A.~Slosar and P.~McDonald,
  JCAP {\bf 0610} (2006) 014
  [arXiv:astro-ph/0604335].

\bibitem{Zhang:2007zzh}
  L.~Zhang, X.~Chen, M.~Kamionkowski, Z.~g.~Si and Z.~Zheng,
  Phys.\ Rev.\  D {\bf 76} (2007) 061301
  [arXiv:0704.2444 [astro-ph]].

\bibitem{Chluba:2008aw}
  J.~Chluba and R.~A.~Sunyaev,
  A \& A 501, 29-47 (2009)
  arXiv:0803.3584 [astro-ph].

\bibitem{Fischler:2010xz}
  W.~Fischler and J.~Meyers,
  arXiv:1011.3501 [astro-ph.CO].

\bibitem{Bashinsky:2003tk}
  S.~Bashinsky and U.~Seljak,
  Phys.\ Rev.\  D {\bf 69} (2004) 083002
  [arXiv:astro-ph/0310198].

\bibitem{Ferreira:1997hj}
  P.~G.~Ferreira and M.~Joyce,
  Phys.\ Rev.\  D {\bf 58} (1998) 023503
  [arXiv:astro-ph/9711102].

\bibitem{Liddle:1998xm}
  A.~R.~Liddle and R.~J.~Scherrer,
  Phys.\ Rev.\  D {\bf 59} (1999) 023509
  [arXiv:astro-ph/9809272].

\bibitem{Binetruy:1999hy}
  P.~Binetruy, C.~Deffayet, U.~Ellwanger and D.~Langlois,
  Phys.\ Lett.\  B {\bf 477} (2000) 285
  [arXiv:hep-th/9910219].




\bibitem{krauss}
  L.~M.~Krauss, C.~Lunardini and C.~Smith,
  arXiv:1009.4666 [hep-ph].

\bibitem{Aver:2010wq}
  E.~Aver, K.~A.~Olive and E.~D.~Skillman,
  JCAP {\bf 1005} (2010) 003
  [arXiv:1001.5218 [astro-ph.CO]].

\bibitem{it}
  Y.~I.~Izotov and T.~X.~Thuan,
  Astrophys.\ J.\  {\bf 710}, L67 (2010)
  [arXiv:1001.4440 [astro-ph.CO]].


\bibitem{steigman}
  G.~Steigman,
  JCAP {\bf 1004 } (2010)  029.
  [arXiv:1002.3604 [astro-ph.CO]].
  
\bibitem{holanda}
  P.~C.~de Holanda and A.~Y.~Smirnov,
  arXiv:1012.5627 [hep-ph].

\bibitem{akhmedov}
  E.~Akhmedov and T.~Schwetz,
  JHEP {\bf 1010} (2010) 115
  [arXiv:1007.4171 [hep-ph]].

\bibitem{giunti}
  C.~Giunti and M.~Laveder,
  Phys.\ Rev.\  D {\bf 82} (2010) 053005
  [arXiv:1005.4599 [hep-ph]].


\bibitem{Doran:2006kp}
  M.~Doran and G.~Robbers,
  JCAP {\bf 0606} (2006) 026
  [arXiv:astro-ph/0601544].

\bibitem{Linder:2008nq}
  E.~V.~Linder and G.~Robbers,
  JCAP {\bf 0806} (2008) 004
  [arXiv:0803.2877 [astro-ph]].

\bibitem{Calabrese:2010uf}
  E.~Calabrese, R.~de Putter, D.~Huterer, E.~V.~Linder and A.~Melchiorri,
  Phys.\ Rev.\  D {\bf 83} (2011) 023011
  [arXiv:1010.5612 [astro-ph.CO]].

\bibitem{dePutter:2010vy}
  R.~de Putter, D.~Huterer and E.~V.~Linder,
  Phys.\ Rev.\  D {\bf 81} (2010) 103513
  [arXiv:1002.1311 [astro-ph.CO]].

\bibitem{Alam:2010tt}
  U.~Alam, Z.~Lukic and S.~Bhattacharya,
  arXiv:1004.0437 [astro-ph.CO].

\bibitem{Hollenstein:2009ph}
  L.~Hollenstein, D.~Sapone, R.~Crittenden and B.~M.~Schaefer,
  JCAP {\bf 0904} (2009) 012
  [arXiv:0902.1494 [astro-ph.CO]].

\bibitem{Xia:2009ys}
  J.~Q.~Xia and M.~Viel,
  JCAP {\bf 0904} (2009) 002
  [arXiv:0901.0605 [astro-ph.CO]].

\bibitem{Giusarma:2011ex}
  E.~Giusarma, M.~Corsi, M.~Archidiacono, R.~de Putter, A.~Melchiorri, O.~Mena and S.~Pandolfi,
  arXiv:1102.4774 [astro-ph.CO].

\bibitem{Galli:2008hm}
  S.~Galli, R.~Bean, A.~Melchiorri and J.~Silk,
  Phys.\ Rev.\  D {\bf 78} (2008) 063532
  [arXiv:0807.1420 [astro-ph]].

\bibitem{Scherrer:2005je}
  R.~J.~Scherrer,
  Phys.\ Rev.\  D {\bf 73} (2006) 043502
  [arXiv:astro-ph/0509890].

\bibitem{Linder:2008ya}
  E.~V.~Linder and R.~J.~Scherrer,
  Phys.\ Rev.\  D {\bf 80} (2009) 023008
  [arXiv:0811.2797 [astro-ph]].

\bibitem{Scherrer:2004au}
  R.~J.~Scherrer,
  Phys.\ Rev.\ Lett.\  {\bf 93} (2004) 011301
  [arXiv:astro-ph/0402316].

\bibitem{camb}
  A.~Lewis, A.~Challinor and A.~Lasenby,
  Astrophys.\ J.\  {\bf 538} (2000) 473
  [arXiv:astro-ph/9911177].


\bibitem{red}
 B.~A.~Reid {\it et al.},
  Mon.\ Not.\ Roy.\ Astron.\ Soc.\  {\bf 404} (2010) 60
  [arXiv:0907.1659 [astro-ph.CO]].


\bibitem{hst}
  A.~G.~Riess {\it et al.},
  Astrophys.\ J.\  {\bf 699}, 539 (2009)
  [arXiv:0905.0695 [astro-ph.CO]].


\bibitem{Lewis:2002ah}
A. Lewis and S. Bridle,
Phys.\ Rev.\ D {\bf 66}, 103511 (2002) (Available from
\texttt{http://cosmologist.info}.)

\bibitem{iocco}
  O.~Pisanti, A.~Cirillo, S.~Esposito, F.~Iocco, G.~Mangano, G.~Miele and P.~D.~Serpico,
  Comput.\ Phys.\ Commun.\  {\bf 178}, 956 (2008)
  [arXiv:0705.0290 [astro-ph]].
  
\bibitem{DeBernardis:2008tk}
  F.~De Bernardis, R.~Bean, S.~Galli, A.~Melchiorri, J.~I.~Silk and L.~Verde,
  Phys.\ Rev.\  D {\bf 79} (2009) 043503
  [arXiv:0812.3557 [astro-ph]].

\bibitem{planck}
    [Planck Collaboration],
  arXiv:astro-ph/0604069.

\bibitem{fishcmb}
J.~R.~Bond, G.~Efstathiou and M.~Tegmark,
 Mon.\ Not.\ Roy.\ Astron.\ Soc.\  {\bf 291} (1997) L33
 [arXiv:astro-ph/9702100].


\end{thebibliography}
\end{document}